\documentclass[12pt]{article}
\usepackage{times}
\usepackage{amsmath}
\usepackage{enumitem}
\usepackage{amsmath}
\usepackage{amssymb}
\usepackage{amsfonts}
\usepackage{hyperref}
\usepackage{mathrsfs}
\usepackage{lineno}
\usepackage{authblk}
\usepackage{graphicx} 
\usepackage{subfig}
\usepackage{float}
\usepackage{xcolor}
\usepackage{comment}

\title{Long-term impact of PM$_{2.5}$ on mortality is exacerbated when wildfire events occur}

\author[1,2]{Federica Spoto\thanks{To whom correspondence should be addressed; E-mail: fspoto@hsph.harvard.edu}}
\author[1]{Francesca Dominici}
\author[3,4]{Tarik Benmarhnia}
\author[1,5]{Danielle Braun}
\author[6,7]{Joan A. Casey}

\affil[1]{Department of Biostatistics, Harvard T.H. Chan School of Public Health, Boston, MA, USA}
\affil[2]{Department of Environmental Health, Harvard T.H. Chan School of Public Health, Boston, MA, USA}
\affil[3]{Scripps Institution of Oceanography, UC San Diego, San Diego, CA, USA}
\affil[4]{Irset Institut de Recherche en Santé, Environnement et Travail, Inserm, University of Rennes, EHESP, Rennes, France }
\affil[5]{Department of Data Science, Dana Farber Cancer Institute, Boston, MA, USA}
\affil[6]{Department of Environmental and Occupational Health Sciences, University of Washington School of Public Health, Seattle, WA, USA}
\affil[7]{Department of Epidemiology, University of Washington School of Public Health, Seattle, WA, USA}

\date{}

\begin{document} 

\baselineskip22pt

\maketitle 

\begin{abstract}
There is extensive evidence that long-term exposure to all-source PM$_{2.5}$ increases mortality. However, to date, no study has evaluated whether this effect is exacerbated in the presence of wildfire events. Here, we study 60+ million older US adults and find that wildfire events increase the harmful effects of long-term all-source PM$_{2.5}$ exposure on mortality, providing a new and realistic conceptualization of wildfire health risks.
\end{abstract}

\section{Main}

There is extensive and consistent evidence that long-term exposure to PM$_{2.5}$ from all sources has an adverse effect on human health and increases the risk of mortality \cite{josey2023air, wu2020evaluating, di2017air, moon2024long, brunekreef2021mortality, pun2017long}. This robust evidence base led to a recent revision of the annual standard for the US National Ambient Air Quality Standards (NAAQS) for PM$_{2.5}$, reducing the allowable threshold from 12 $\mu g/m^3$ to 9 $\mu g/m^3$. These more stringent standards are expected to yield significant public health benefits, including the prevention of up to 4,500 premature deaths per year across the US. Moreover, the net economic benefits—accounting for reduced healthcare expenditures, fewer workdays lost, and other health-related gains minus the estimated costs of regulatory compliance—are projected to reach approximately \$46 billion by 2032 \cite{EPAfinal}. \\

The increasing frequency and geographic scope of wildfires and their smoke have increased interest in their health impacts. Climate-induced increases in fuel aridity, historical fire suppression, and the growing human presence in wildland areas have driven increased wildfire activity.\cite{parks2020warmer} Since 2000, an average of 70,025 wildfires burned 7.0 million acres annually in the US, more than twice the average of the 1990s \cite{epa_wildfires, nifc_statistics}. Beginning in 2016, these events have reversed decades of progress in reducing PM$_{2.5}$ concentrations in nearly three-quarters of US states \cite{Burke2023}. 

Several studies have reported an association between short- and medium-term exposure to wildfire events and higher rates of emergency department visits, hospitalizations, and consultations for respiratory and cardiovascular problems \cite{Chen2021, Groot2019, To2021, Reid2024}, as well as a higher risk of all-cause mortality \cite{Chen2021mortality, Ye2022, Doubleday2020} and and deaths due to COVID-19 \cite{Zhou2021}. In these studies, wildfire events were defined either based on the levels of wildfire PM$_{2.5}$--fine particulate matter originating from wildfire smoke--or all-source PM$_{2.5}$ during periods when wildfires burned. Recent studies have documented associations between long-term exposure to wildfire PM$_{2.5}$ and adverse respiratory, mental, and overall health outcomes, including an increased risk of mortality \cite{Connolly2024, Ma2024, Wei2023}. \\ 

Despite extensive research, we lack evidence on whether the risk conferred by long-term exposure to all-source PM$_{2.5}$ becomes amplified when wildfire events occur. Our study addresses the research question "Do wildfire events exacerbate the adverse mortality effects associated with all-source PM$_{2.5}$?" 

\begin{figure}[ht!]
\centering
\subfloat{
\includegraphics[width=\linewidth]{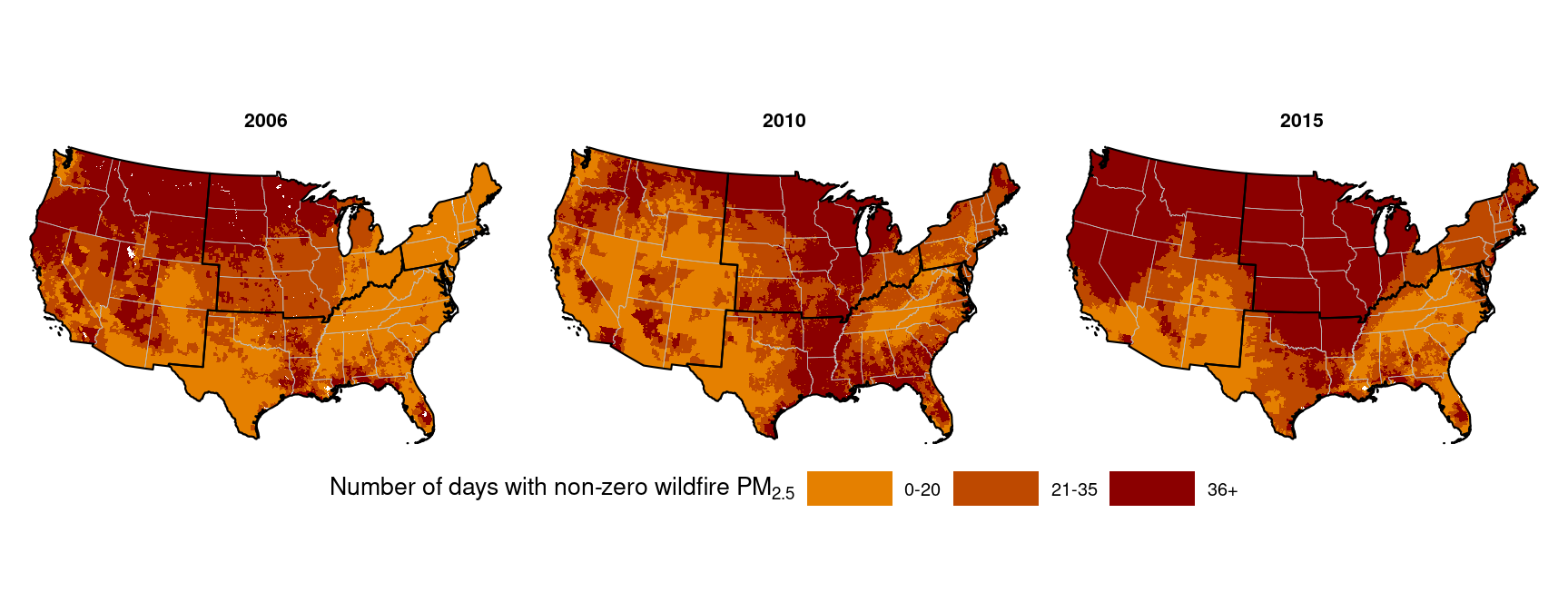}} 
\caption{\textbf{Wildfire event distribution.} The plot represents the distribution of the three wildfire event day strata over the contiguous US for the years 2006, 2010, and 2015. The three wildfire event day strata are (1) 0 to 20 days with non-zero wildfire PM$_{2.5}$ per year, (2) 21 to 35 days with non-zero wildfire-PM$_{2.5}$ per year, and (3) more than 35 days with non-zero wildfire PM$_{2.5}$ per year.}
\label{fig:fig1}
\end{figure}

Our study cohort comprises more than 60 million US adults aged 65 years and older, with a total of 17,608,624 recorded deaths from 2007 to 2016. Exposure to PM$_{2.5}$ is characterized as the annual average concentration of all-source PM$_{2.5}$ between 2006 and 2015, as utilized by Wu et al. \cite{wu2020evaluating}, derived from the daily estimates developed by Di et al. \cite{DI2019104909}. 
The potential effect modification by wildfire events is defined 
as the "number of days with non-zero wildfire PM$_{2.5}$ per year" \cite{Casey2024}. Estimates of daily wildfire PM$_{2.5}$ are obtained from a previously validated daily model for the contiguous US over the period 2006--2015 \cite{childs2022daily}. This metric effectively captures both the duration and frequency of exposure to wildfire events in each year. 
For our analyses, we categorize this effect modifier into three distinct strata based on annual counts of wildfire event days: (1) 0 to 20 non-zero wildfire PM$_{2.5}$ days per year (approximating the first tertile); (2) 21 to 35 non-zero wildfire PM$_{2.5}$ days per year (approximating the second tertile); and (3) more than 35 non-zero wildfire PM$_{2.5}$ days per year. Figure \ref{fig:fig1} illustrates the spatial and temporal distribution of these strata across the US. 
Exposure and effect modification variables are aggregated at the ZIP code level and linked to Medicare beneficiaries according to their residential ZIP code and year of assessment. This linkage enables us to associate mortality data with exposure measures and effect modifiers from the previous year.

Within each of the three wildfire event days strata, we estimate the association between annual all-source PM$_{2.5}$ levels and all-cause mortality counts in the subsequent year for each ZIP code using Poisson regression models allowing for a nonlinear exposure–response relationship between all-source PM$_{2.5}$ and mortality. These models are 
also stratified by individual-level characteristics and follow-up year. To control for additional potential confounding, we adjust for 10 ZIP code or county-level covariates, four ZIP code–level meteorological variables, and indicators for geographic region (Northeast, South, Midwest, West) and calendar year  \cite{wu2020evaluating}. To examine additional potential effect modification by geography or socioeconomic status, we further stratify the cohort by region and ZIP code-level poverty, using 2010 US Census data. ZIP codes are categorized as high or low poverty based on whether 15\% or more of the population lives below the federal poverty level. Geographic and poverty-related modifiers, along with potential confounders, were assigned for the year corresponding to the outcome.

Our main parameter of interest is the relative change in all-cause mortality risk associated with deviations from the current National Ambient Air Quality Standard (NAAQS) for annual PM$_{2.5}$ levels (9 $\mu g/m^3$), under each strata (0–20, 21–35, and $>$35 wildfire PM$_{2.5}$ days per year).

\begin{figure}[ht!]
    \centering
    \includegraphics[width=\linewidth]{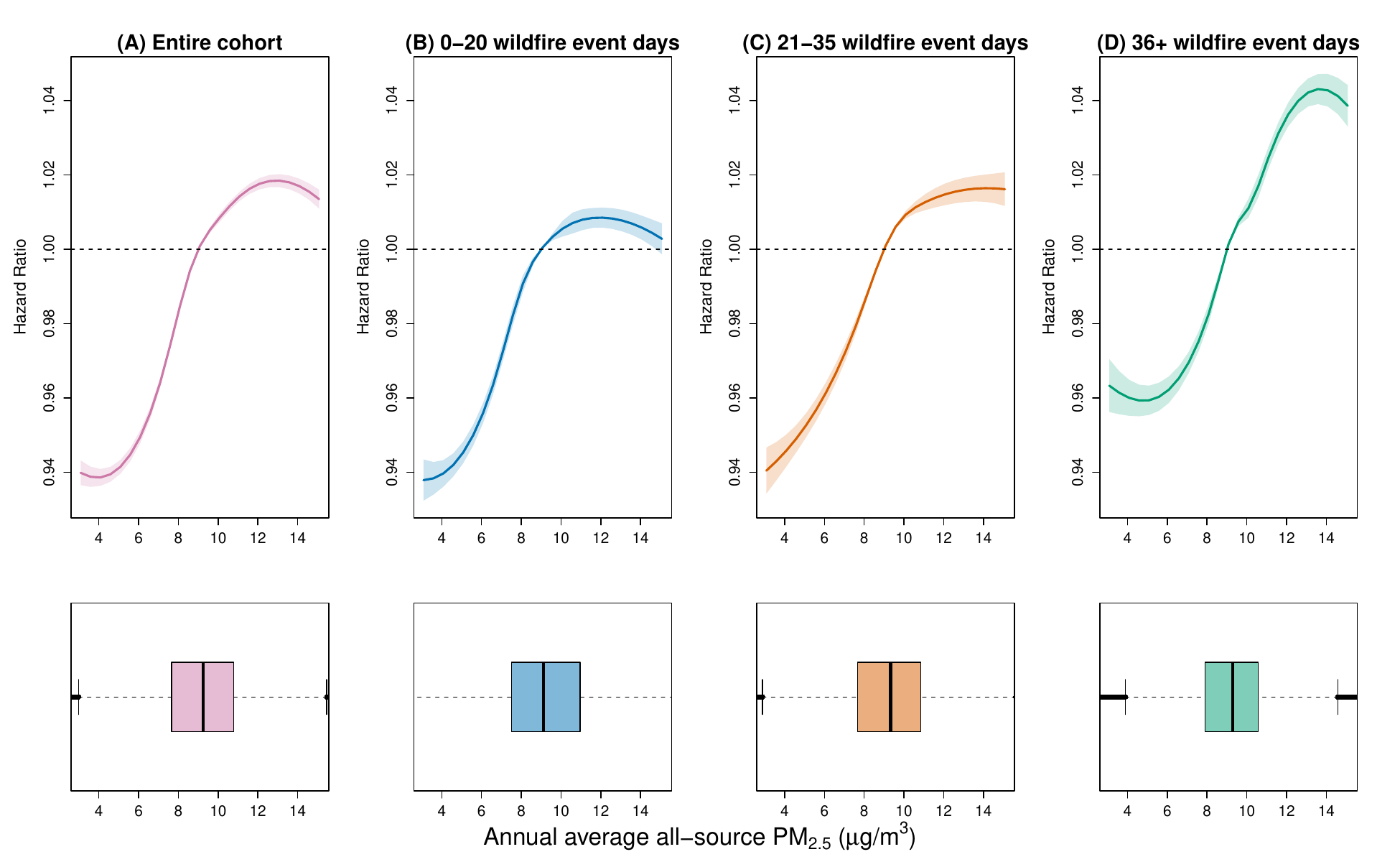}
    \caption{\textbf{Mortality hazard ratio with 95\% confidence intervals and all-source PM$_{2.5}$ distribution.} The plots illustrate the exposure-response curve and distribution of all-source PM$_{2.5}$ values, ranging from the $1^{st}$ to the $99^{st}$ percentile of the entire cohort all-source PM$_{2.5}$ distribution. The top plots represent the hazard ratio of mortality compared to the current NAAQS for annual all-source PM$_{2.5}$(9 $\mu g/m^3$) for the entire cohort and stratifying by the three wildfire days strata. The lower plots show the distribution of all-source PM$_{2.5}$ for the entire cohort, as well as stratified by the three wildfire day strata.}
    \label{fig:all_us}
\end{figure}

During the study period, the mean (SD) level of annual average all-source PM$_{2.5}$ exposure was 9.0 $\mu g/m^3$ (2.7 $\mu g/m^3$) and wildfire PM$_{2.5}$ was 0.37 $\mu g/m^3$ (0.34 $\mu g/m^3$). Table \ref{tab:tab1} describes the cohort characteristics across the contiguous US during the study period.\\

We find that wildfire event days modify the relationship between all-source PM$_{2.5}$ and mortality (Figure \ref{fig:all_us}). Being exposed to a higher number of days with non-zero wildfire PM$_{2.5}$ during the year was associated with higher mortality risk in the following year, with the highest risk for those exposed to 12 $\mu g/m^3$ of all-source PM$_{2.5}$ and experiencing more than 35 days with non-zero wildfire PM$_{2.5}$ days per year. \\

\begin{figure}[ht!]
    \centering
    \includegraphics[width=0.9\linewidth]{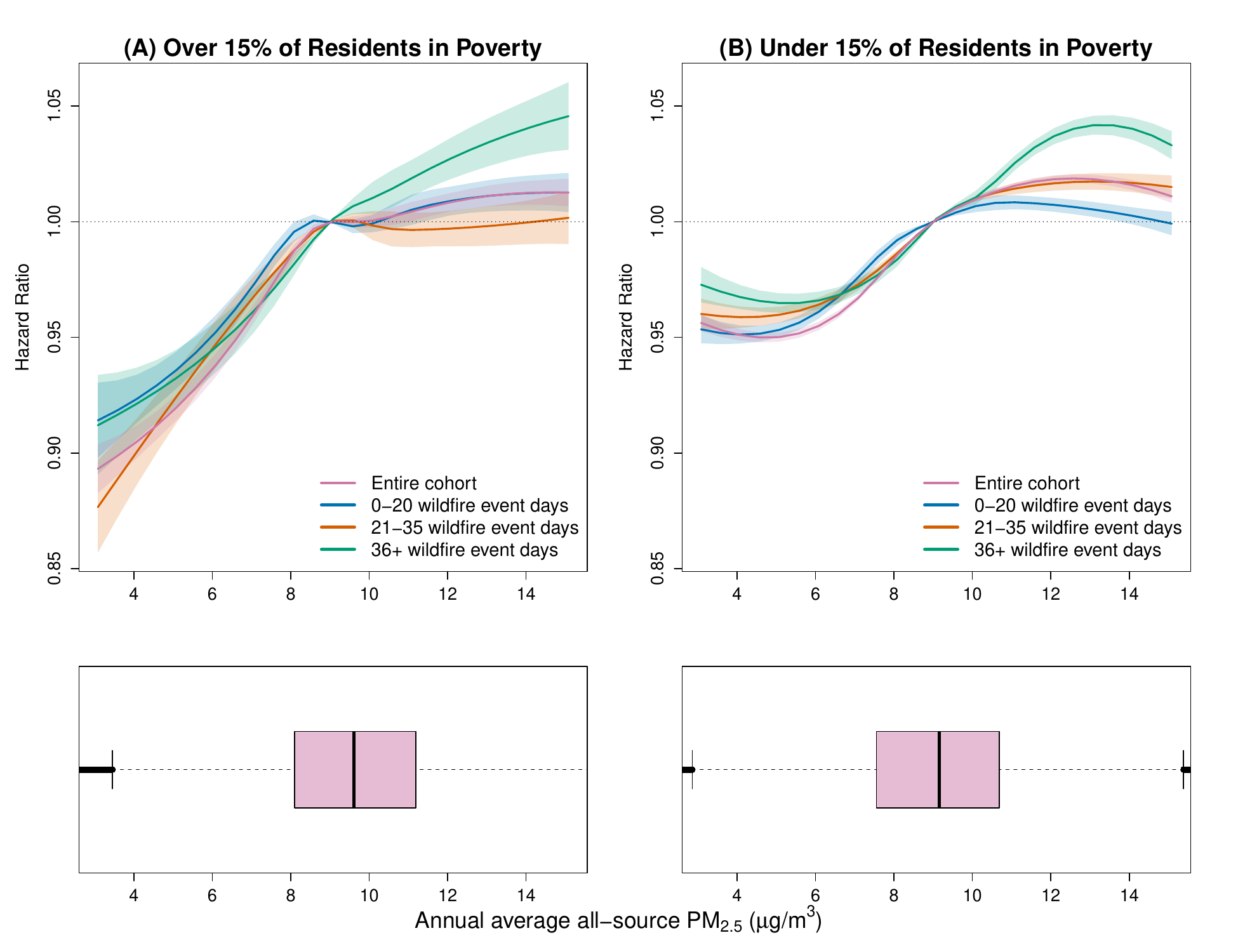}
    \caption{\textbf{Mortality hazard ratio with 95\% confidence intervals and all-source PM$_{2.5}$ distribution stratifying by poverty level.} The plots illustrate the exposure-response curve and distribution of all-source PM$_{2.5}$ values, ranging from the $1^{st}$ to the $99^{st}$ percentile of the entire cohort all-source PM$_{2.5}$ distribution. The top plots represent the hazard ratio of mortality compared to the current NAAQS for annual all-source PM$_{2.5}$(9 $\mu g/m^3$) stratifying by poverty level. The lower plots represent the distribution of all-source PM$_{2.5}$ for the two poverty-based strata. }
    \label{fig:poverty}
\end{figure}

When further stratifying the cohort by ZIP code poverty levels, we find that the all-source PM$_{2.5}$-mortality association is further modified, especially at lower levels of annual all-source PM$_{2.5}$ exposure where the exposure–response relationship is steeper in higher-poverty ZIP codes (Figure \ref{fig:poverty}). This finding indicates that incremental increases in PM$_{2.5}$ concentrations at relatively low baselines may lead to disproportionately higher mortality risks in socioeconomically disadvantaged communities. At higher levels of all-source PM$_{2.5}$, we find different risk patterns based on area-level poverty. In lower-poverty ZIP codes, increases in mortality risk are more pronounced with higher numbers of wildfire events. In contrast, in ZIP codes where the poverty rate exceeds 15\% we observe a different pattern. The mortality risks associated with all-source PM$_{2.5}$ exposure modified by wildfire event are comparable for areas experiencing 0–20 and 21–35 wildfire event days annually, with risk curves largely overlapping. However, a marked increase in risk emerges for areas experiencing more than 35 wildfire event days per year, and risk does not appear to plateau at high all-source PM$_{2.5}$ as in other strata and in the lower-poverty ZIP codes. \\

We then evaluated region differences by stratifying the overall all-source PM$_{2.5}$-mortality association by census region. We observed varying shapes of the exposure-response curve by census region without considering wildfire event day effect modification. When adding the number of wildfire days per year as an effect modifier, the results remained variable by region (Figure \ref{fig:region}).  \\

In this study, we find that experiencing more days of non-zero wildfire PM$_{2.5}$ amplifies the association between long-term exposure to all-source PM$_{2.5}$ and all-cause mortality in a cohort of $>$60 million older US adults, providing a new and realistic conceptualization of wildfire health risks.

\appendix
\section{Online Methods}

\renewcommand{\thefigure}{S\arabic{figure}} 
\renewcommand{\thetable}{S\arabic{table}}   
\setcounter{figure}{0} 
\setcounter{table}{0} 

\subsection{Study population}  

Our study cohort comprises 60,999,431 Medicare beneficiaries 65 years and older for the period 2007--2016 in the contiguous US. Medicare claims data is obtained from the Centers for Medicare and Medicaid Services and includes individual-level data on age, sex, race/ethnicity, date of death, and residential ZIP code. We use residential ZIP codes to link exposure, effect modifiers, and area-level covariate information for each patient. 
Each Medicare beneficiary is assigned a unique patient identifier, which allows for longitudinal tracking. Cohort inclusion began in 2007 for those already enrolled prior to January 1, 2007 or upon their enrollment after 2007. Each enrollee is monitored annually until death or the end of the study period on December 31, 2016. 

\subsection{Exposure assessment}

For the period 2006-2015, we consider the annual average of all-source PM$_{2.5}$ concentrations at the ZIP code level as the exposure 
derived from daily estimates by Di et al. \cite{DI2019104909}. These daily estimates 
were obtained from an ensemble prediction model \cite{DI2019104909}. We then obtained ZIP code level estimates by averaging grid cells with centroids inside each ZIP code. For P.O. Box-only ZIP codes, all-source PM$_{2.5}$ concentrations were assigned based on the nearest grid cell. Finally, annual concentrations were obtained by averaging the daily levels.
We assign exposure based on the Medicare enrollee's residential ZIP code in each calendar year, linking the prior year's exposure to the current year's mortality outcome. \\

\subsection{Potential effect modifiers}

For the period 2006-2015, we consider the number of days with non-zero wildfire PM$_{2.5}$ per year as a potential effect modifier of the all-source PM$_{2.5}$-mortality association.
We obtain daily levels of wildfire PM$_{2.5}$ at 10-km$^2$ grid resolution from Childs et al. \cite{childs2022daily}. We overlay the 10-km$^2$ grid and the ZIP code boundaries and use area-weighting to obtain daily ZIP code level wildfire PM$_{2.5}$ estimates. For P.O. Box-only ZIP codes, wildfire PM$_{2.5}$ concentrations are assigned based on the nearest non-P.O. Box ZIP code.  We then use the daily ZIP code estimates of wildfire PM$_{2.5}$ to compute the annual number of non-zero wildfire PM$_{2.5}$ days for each ZIP code. Following \cite{Casey2024}, we count the number of days where wildfire PM$_{2.5}$ concentrations exceed zero per year in each ZIP code. Lastly, we categorize the annual number of non-zero wildfire PM$_{2.5}$ days in three strata: (1) 0 to 20 non-zero wildfire PM$_{2.5}$ days per year (approximating the first tertile); (2) 21 to 35 non-zero wildfire PM$_{2.5}$ days per year (approximating the second tertile); and (3) more than 35 non-zero wildfire PM$_{2.5}$ days per year. We assign wildfire event days based on the Medicare enrollee's residential ZIP code in each calendar year, linking the prior year's number of days with non-zero wildfire PM$_{2.5}$ to the current year's mortality. \\

In addition, we examined two additional potential effect modifiers: ZIP code–level poverty and US Census region. ZIP code–level poverty was dichotomized, classifying participants based on whether they resided in ZIP codes where 15\% or more of the population lived below the poverty threshold, versus those in areas with less than 15\% of residents living in poverty. The US Census region was categorized into four geographically distinct strata: Northeast, Midwest, South, and West. Geographic and poverty-related modifiers, along with potential confounders, were assigned for the corresponding year of the mortality outcome.

\subsection{Potential confounders}  

In order to address confounding bias from community-level factors, we use information from various sources, including ZIP code-level socioeconomic status (SES) indicators from the 2000 and 2010 Census and the 2005--2012 American Community Surveys (ACS) and county-level information from the Centers for Disease Control and Prevention’s Behavioral Risk Factor Surveillance System (BRFSS). We spatially align all data with postal ZIP codes to match Medicare resolution. This includes: (i) two county-level variables: average body mass index and smoking rate; (ii) eight ZIP census variables: proportion of Hispanic and black residents, median household income, median home value, proportion of residents in poverty, proportion of residents with a high school diploma, population density, and proportion of residents that own their house; and (iii) four ZIP code-level meteorological variables from Gridmet \cite{abatzoglou2013development,gorelick2017google}: summer (June to September) and winter (December to February) average maximum daily temperature and relative humidity. The ZIP code–level meteorological variables are obtained using area-weighted aggregations based on daily temperature and humidity data on 4-km$^2$ gridded rasters from Gridmet via Google Earth Engine. 
Finally, we create categorical variables for the four geographic US census regions (Northeast, South, Midwest, and West), and calendar year (2007--2016), to help account for any residual or unmeasured spatial and temporal confounding. This set of confounders corresponds to the ones used in \cite{wu2020evaluating}. 

\subsection{Data linkage}  
Mortality data are available at the ZIP code level. 
ZIP codes that have incomplete data or no Medicare patients are excluded from the analysis (n = 9,882). The analysis includes 34,444 ZIP codes ($\sim$80\% of US ZIP codes) with complete data on outcomes, exposures, effect modifiers, and confounders. 

\subsection{Statistical analysis}  \label{sec:stat_an}
We conduct an analysis to examine whether the number of wildfire event days serves as an effect modifier of the all-source PM$_{2.5}$--mortality association. 

To assess this potential effect modification, we implement a stratified Poisson regression approach, where the study cohort is divided into three strata based on the number of days with $>$0 $\mu g/m^3$ of wildfire PM$_{2.5}$ per year: (1) 0–20 days, (2) 21–35 days, and (3) more than 35 days. Within each wildfire day stratum, we independently estimate the exposure–response relationship between annual average all-source PM$_{2.5}$ levels and mortality. This stratified design allows us to directly compare the hazard ratios for PM$_{2.5}$ across different levels of wildfire days.

The Poisson model is specified for each stratum using ZIP code–level death counts by calendar year and follow-up year as the outcome, with annual average all-source PM$_{2.5}$ as a time-varying exposure and the log of person-years included as an offset term. We incorporate natural spline terms to model non-linear associations with PM$_{2.5}$. Models are adjusted for 10 ZIP code or county-level socioeconomic and demographic covariates, four ZIP code–level meteorological variables, as well as fixed effects for geographic region (Northeast, South, Midwest, West) and calendar year to account for geographic and temporal heterogeneity. To control for individual-level variability in baseline mortality risk, each model is stratified by four individual-level characteristics: (i) a 5-year category of age at entry (65 to 69, 70 to 74, 75 to 79, 80 to 84, 85 to 89, 90 to 94, 95 to 99, and above 100 years of age), (ii) race/ethnicity (Hispanic, Native American and Alaska Native, Non-Hispanic Asian, Non-Hispanic Black, Non-Hispanic White, and other [e.g., multi-racial and missing]), (iii) sex (male or female), and (iv) an indicator variable for Medicaid eligibility (a surrogate for individual-level SES). For more details, see \cite{wu2020evaluating}.

For each wildfire strata, the Poisson model is expressed as:
\begin{align*}
    \log \mathbb{E}[\text{death counts}] & \sim \text{ns(all-source PM}_{2.5}\text{, df)} \\&  + \text{area-level risk factors} + \text{meteorological variables}\\ 
    & + \text{year} + \text{region} + \text{offset(log[person-years])} \\
    & + \text{strata(age, race, sex, Medicaid status, follow-up year)}.
\end{align*}

The degrees of freedom (df) for the natural spline terms in each model were selected independently, using the Akaike Information Criterion (AIC). The df that minimized the AIC was chosen for each wildfire stratum. Table \ref{tab:df_aic} summarizes the optimal degrees of freedom selected in the Poisson models.

In addition, we consider two additional potential effect modifiers: ZIP code–level poverty and US census region. ZIP code–level poverty is operationalized as a binary variable, dividing the cohort into two groups: those living in ZIP codes where 15\% or more of residents live below the federal poverty level, and those living in ZIP codes with less than 15\% living in poverty. Census region is categorized into four geographic strata: Northeast, Midwest, South, and West.

We apply the previously defined Poisson regression model across all three wildfire exposure strata within each level of these secondary effect modifiers. That is, the model is fit separately for each wildfire exposure group within each poverty group and each census region. When stratifying by poverty or region, we exclude poverty or region as a covariate from the corresponding models to avoid redundancy and ensure appropriate interpretation of the stratum-specific estimates.

Similar to the main model, the degrees of freedom for the natural spline terms in each model were determined independently. This was achieved by minimizing the Akaike Information Criterion (AIC) for each combination of wildfire category and poverty stratum, as well as for each combination of wildfire category and region. The optimal degrees of freedom selected for the Poisson models in the stratified analyses are presented in Table \ref{tab:df_aic}.

Effect estimates are presented as hazard ratios comparing mortality risk at varying levels of annual average all-source PM$_{2.5}$ to the current US National Ambient Air Quality Standard (NAAQS) of 9 $\mu g/m^3$. These hazard ratios are estimated separately within each wildfire exposure stratum, providing insight into whether the mortality risk per unit PM$_{2.5}$ varies depending on wildfire day levels.

\section{Code availability}
R code implementing the analyses can be found at \url{https://github.com/NSAPH-Projects/totalPM-smokePM-mortality}.

\section*{Acknowledgments}
JA Casey received funding from the National Institute on Aging (R01AG071024).

\bibliographystyle{unsrt}
\bibliography{biblio}
\newpage

\section*{Supplementary figures and tables}

\begin{table}[ht!]
    \centering
    {\footnotesize
    \begin{tabular}{l|c}
        Variables & Medicare beneficiaries  \\
        \hline
         Number of individuals & 60,999,431 \\
         Number of deaths & 17,608,624 (28.9\%) \\
         Total person years & 391,702,092 \\
         Median years of follow-up & 9 \\
         \hline
         Individual-level characteristics & \\
         \hline
         Age at entry (years) & \\
         \hspace{0.2cm} 65–74 (\%) & 88.9\\
         \hspace{0.2cm} 75–84 (\%) & 9.4\\
         \hspace{0.2cm} 85–94 (\%) & 1.5\\
         \hspace{0.2cm} 95 or above (\%) & 0.2\\
         Sex & \\
         \hspace{0.2cm} Female (\%) & 55.4\\
         Race and ethnicity & \\
         \hspace{0.2cm} Hispanic (\%) & 2.1\\
         \hspace{0.2cm} Native American and Alaska Native (\%) & 0.3 \\
         \hspace{0.2cm} Non-Hispanic Asian (\%) & 2.1 \\
         \hspace{0.2cm} Non-Hispanic Black (\%) & 9.0 \\
         \hspace{0.2cm} Non-Hispanic White (\%) & 83.4 \\
         Medicaid eligibility & \\
         \hspace{0.2cm} Eligible (\%) & 12.9 \\
         Region of residence & \\
         \hspace{0.2cm} Northeast (\%) & 20.15 \\
         \hspace{0.2cm} Midwest (\%) & 22.15 \\
         \hspace{0.2cm} West (\%) & 20.67 \\
         \hspace{0.2cm} South (\%) & 37.03 \\
         \hline
         Area-level risk factors	& \\
         \hline
         Ever smoked (\%) & 46.7 \\
         Below poverty level (\%) & 10.1\\
         Less than high school education (\%) & 23.5\\
         Owner-occupied housing (\%) & 71.6\\
         Hispanic (\%) & 9.5\\
         Non-Hispanic Black (\%) & 8.9\\
         Population density (persons/km$^2$) & 1556.0 (5160.5) \\
         Mean BMI (kg/m$^2$) & 27.80 (1.02) \\
         Median household income (\$1000) & 53.1 (23.0)\\
         Median home value (\$1000) & 187.7 (157.8) \\
         Meteorological variables, mean (SD) & \\	 
        \hspace{0.2cm} Summer temperature (°C)	& 29.7 (3.8) \\
        \hspace{0.2cm} Winter temperature (°C)	& 7.4 (7.3)\\
        \hspace{0.2cm} Summer relative humidity (\%) & 86.8 (11.8) \\
        \hspace{0.2cm} Winter relative humidity (\%) & 85.6 (7.2) \\
        \hspace{0.2cm} All-source PM$_{2.5}$ concentration ($\mu g/m^3$) & 9.0 (2.7) \\
        \hspace{0.2cm} Wildfire PM$_{2.5}$ concentration ($\mu g/m^3$) & 0.4 (0.3)\\
        \hspace{0.2cm} Non-zero wildfire PM$_{2.5}$ days per year & 31.8 (21.7)
    \end{tabular}}
    \caption{\textbf{Characteristics for the Medicare study cohort, 2006--2016.} Mortality and individual-level characteristics were obtained from the Centers for Medicare and Medicaid Services (CMS), and ZIP code socioeconomic status (SES) was obtained from the 2000 and 2010 Census and the 2005–2012 American Community Surveys (ACS) and county-level behavioral risk factor variables were obtained from the Centers for Disease Control and Prevention. All-source PM$_{2.5}$ and wildfire PM$_{2.5}$ concentrations were obtained respectively from \cite{di2017air} and \cite{childs2022daily}. Meteorological variables were obtained from Gridmet via Google Earth Engine. }
    \label{tab:tab1}
\end{table}

\begin{table}[ht!]
    \centering
    \begin{tabular}{c|ccccccc}
        \textbf{Stratification} & \textbf{Main} & \multicolumn{2}{c}{\textbf{Poverty}}& \multicolumn{4}{c}{\textbf{Region}} \\
       
        & & $>$15\% & $\le$15\% & Northeast & Midwest & South & West  \\
        \hline
        Entire cohort   & 5 & 6 & 5 & 6 & 6 & 6 & 6 \\
        0–20 days       & 6 & 6 & 6 & 6 & 6 & 6 & 6 \\
        21–35 days      & 4 & 5 & 4 & 5 & 3 & 5 & 6 \\
        36+ days        & 6 & 5 & 6 & 5 & 5 & 6 & 3 \\
    \end{tabular}
    \caption{\textbf{Optimal degrees of freedom for spline terms.} Degrees of freedom selected for the natural spline of annual average all-source PM$_{2.5}$ in each wildfire, poverty and region category stratum. Values were selected to minimize the Akaike Information Criterion (AIC).}
    \label{tab:df_aic}
\end{table}

\begin{figure}[ht!]
    \centering
    \includegraphics[width=\linewidth]{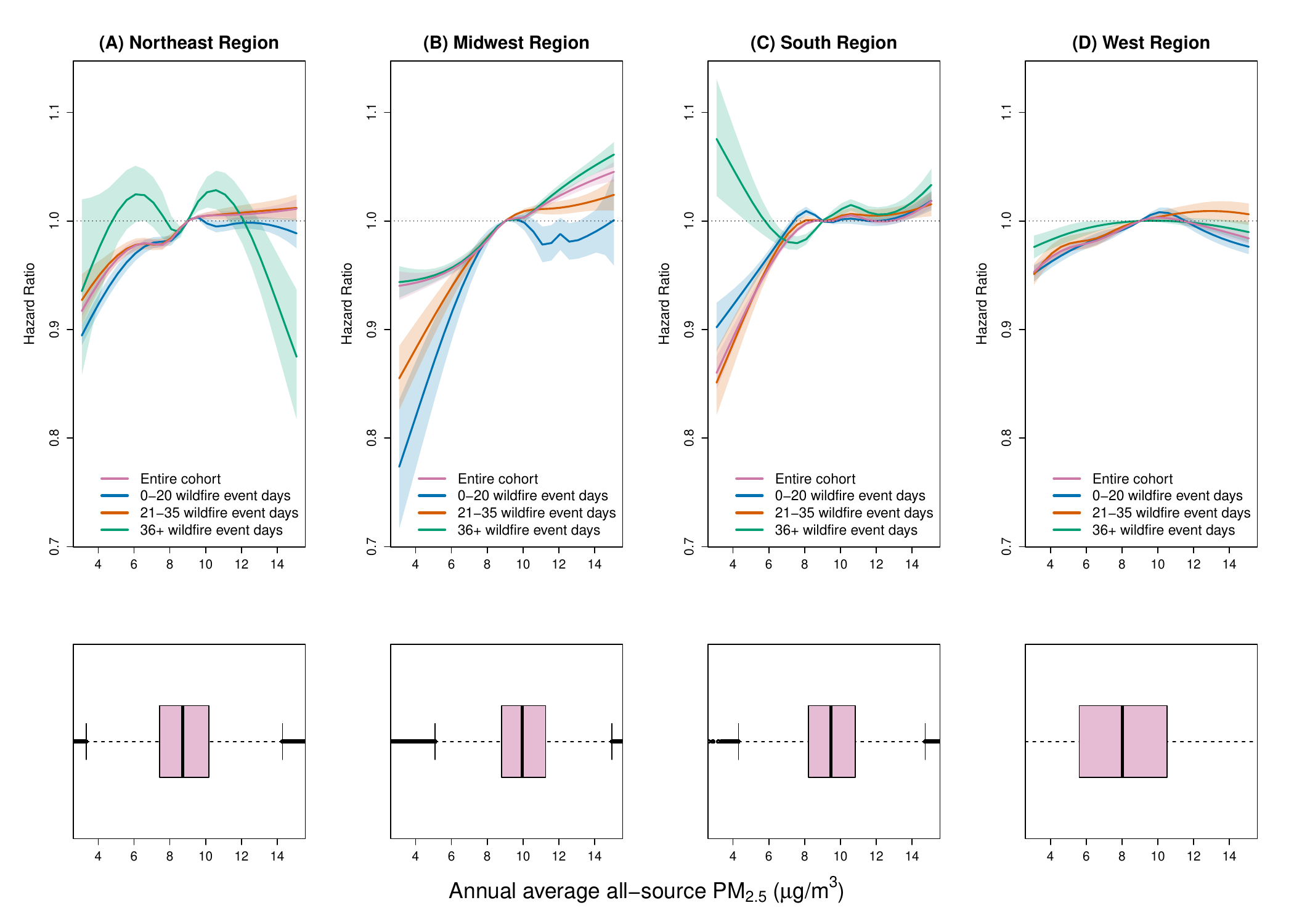}
    \caption{\textbf{Mortality hazard ratio with 95\% confidence intervals and all-source PM$_{2.5}$ distribution stratifying by region.}  The plots illustrate the exposure-response curve and distribution of all-source PM$_{2.5}$ values stratifying by region, ranging from the $1^{st}$ to the $99^{st}$ percentile of the entire cohort all-source PM$_{2.5}$ distribution. The top plots represent the hazard ratio of mortality compared to the current NAAQS for annual all-source PM$_{2.5}$(9 $\mu g/m^3$) stratifying by region. The lower plots represent the distribution of all-source PM$_{2.5}$ for each region. }
    \label{fig:region}
\end{figure}

\end{document}